\documentclass{article}

\usepackage{amssymb,amsmath,eucal}
\usepackage[dvips]{lscape,graphicx}
\usepackage{cite}

\newcommand{\ct}{\cite}
\newcommand{\lb}{\label}

\newcommand{\bc}{\begin{center}}
\newcommand{\ec}{\end{center}}
\newcommand{\bd}{\begin{displaymath}}
\newcommand{\ed}{\end{displaymath}}
\newcommand{\be}{\begin{equation}}
\newcommand{\ee}{\end{equation}}
\newcommand{\ba}{\begin{array}}
\newcommand{\ea}{\end{array}}
\newcommand{\bt}{\begin{tabular}}
\newcommand{\et}{\end{tabular}}
\newcommand{\un}{\underline}
\newcommand{\ov}{\overline}
\newcommand{\bp}{\begin{picture}}
\newcommand{\ep}{\end{picture}}
\newcommand{\bfi}{\begin{figure}}
\newcommand{\efi}{\end{figure}}

\def\fun#1#2{\lower3.6pt\vbox{\baselineskip0pt\lineskip.9pt
\ialign{$\mathsurround=0pt#1\hfil##\hfil$\crcr#2\crcr\sim\crcr}}}

\begin{document}
\vspace{1cm}

\title{\Large \bf [SU(5)]$^{\bf 3}$ SUSY unification}
\author{{\large \bf Larisa Laperashvili, Dmitri Ryzhikh }\\
\it Institute of Theoretical and Experimental Physics,\\
\it Moscow, Russia }

\date{}

\maketitle

\pagenumbering{arabic}

\noindent

\section{Introduction}

The modern physics of electroweak and strong interactions is described by
the Standard Model (SM), unifying the Glashow--Salam--Weinberg electroweak
theory and QCD -- theory of strong interactions.

The gauge group of the SM is:
\begin{equation}
   SMG = SU(3)_c\times SU(2)_L\times U(1)_Y,            \lb{2}
\end{equation}
which describes the present elementary particle physics up to the
scale $\sim 100$ GeV.

Considering the physical processes at very small (Planck scale) distances,
physicists can make attempts to explain the well--known laws of physics
as a consequence of the more fundamental laws of Nature. Random Dynamics (RD)
was suggested and developed in Refs.[\citen{2a}--\citen{2k}] as a theory of physical
processes proceeding at small distances of order of the Planck length
$\lambda_P$:
\begin{equation}
\lambda_P=M_{Pl}^{-1}, \quad{\mbox{where}} \quad
                M_{Pl}=1.22\cdot 10^{19}\,{\mbox{GeV}}.     \lb{1}
\end{equation}
Having an interest in fundamental laws of physics leading to the description
of the low--energy SM phenomena, observed by the contemporary experiment,
we can consider two possibilities:

\vspace{0.1cm}

1. At very small (Planck scale) distances
{\un{\it our space--time is continuous}}, and there exists a theory with
a very high symmetry.

\vspace{0.1cm}

2. At very small distances
{\un{\it our space--time is discrete}}, and this discreteness
influences on the Planck scale physics.

The item 2 is a base of the RD theory.

The theory of Scale Relativity (SR) \ct{10na} is also related with item 2
and has a lot in common with RD. In the SR the resolution of experimental
measurements plays in quantum physics a completely new role with respect
to the classical one and there exists a minimal scale of the space-time
resolution: $\epsilon_{min}=\lambda_P$, which can be considered as a
fundamental scale of our Nature. In this case, our (3+1)--dimensional
space is discrete on the fundamental level. This is an initial point of
view of the present theory, but not an approximation.

The lattice model of gauge theories is the most convenient formalism
for the realization of the RD ideas. In the simplest case we can imagine
our space--time as a regular hypercubic (3+1)--lattice with the parameter
$a$ equal to the fundamental scale:
\begin{equation}
 a = \lambda_P = 1/M_{Pl} \sim 10^{-33}\, cm.                   \lb{3}
\end{equation}
But, in general, we do not know (at least on the level of our today
knowledge) what lattice--like structure plays role in the description
of the physical processes at very small distances.

\section{G-theory, or Anti-Grand Unification Theory (AGUT)}

Most efforts to explain the Standard Model (SM) describing well all
experimental results known today are devoted to Grand Unification
Theories (GUTs). The supersymmetric extension of the SM consists of taking the
SM and adding the corresponding supersymmetric partners \ct{32a}.  The Minimal
Supersymmetric Standard Model (MSSM) shows \ct{33a} the possibility of the
existence of the grand unification point at
$\mu_{GUT}\sim 10^{16}$ GeV.
Unfortunately, at present time experiment does not indicate any manifestation
of the supersymmetry. In this connection, the Anti--Grand Unification
Theory (AGUT) was developed in Refs.[\citen{2a}--\citen{2k}] and [\citen{17p}--\citen{38}]
as a realistic alternative to SUSY GUTs. According to this theory, supersymmetry does
not come into the existence up to the Planck energy scale (\ref{1}).
The Standard Model (SM) is based on the group SMG described by Eq.(\ref{2}).
AGUT suggests that at the scale $\mu_G\sim \mu_{Pl}=M_{Pl}$
there exists the more fundamental group $G$ containing $N_{gen}$
copies of the Standard Model Group SMG:
\begin{equation}
G = SMG_1\times SMG_2\times...\times SMG_{N_{gen}}\equiv (SMG)^{N_{gen}},
                                                  \lb{76y}
\end{equation}
where $N_{gen}$ designates the number of quark and lepton generations.

If $N_{gen}=3$ (as AGUT predicts), then the fundamental gauge group G is:
\begin{equation}
    G = (SMG)^3 = SMG_{1st gen.}\times SMG_{2nd gen.}\times SMG_{3rd gen.},
                                        \lb{77y}
\end{equation}
or the generalized one:
\begin{equation}
         G_f = (SMG)^3\times U(1)_f,           \lb{78y}
\end{equation}
which was suggested by the fitting of fermion masses of the SM
(see Refs.\ct{35}).

Recently a new generalization of AGUT was suggested in Refs.\ct{37}:
\begin{equation}
           G_{\mbox{ext}} = (SMG\times U(1)_{B-L})^3,    \lb{79y}
\end{equation}
which takes into account the see--saw mechanism with right-handed neutrinos,
also gives the reasonable fitting of the SM fermion masses and describes
all neutrino experiments known today.

By reasons considered in this paper, we prefer
not to use the terminology "Anti-grand unification theory, i.e. AGUT",
but call the theory with the group of symmetry $G$, or $G_f$,
or $G_{ext}$, given by Eqs.(\ref{76y})-(\ref{79y}), as "G--theory",
because, as it will be shown below, we have a possibility of the
Grand Unification near the Planck scale using just this theory.

The group $G_f$ contains the following gauge fields:
$3\times 8 = 24$ gluons, $3\times 3 = 9$ W-bosons and $3\times 1 + 1 = 4$
Abelian gauge bosons. The group $G_{ext}$ contains:
$3\times 8 = 24$ gluons, $3\times 3 = 9$ W-bosons and $3\times 1 +
3\times 1 = 6$ Abelian gauge bosons.

There are five Higgs fields in AGUT, extended by Froggatt and Nielsen \ct{35}
with the group of symmetry $G_f$ given by Eq.(\ref{78y}).
These fields break AGUT to the SM what means that their vacuum expectation
values (VEV) are active.
The authors of Refs.\ct{35} used three parameters -- three independent VEVs with
aim to find the best fit to conventional experimental data
for all fermion masses and mixing angles in the SM.
The result was encouraging.

The extended AGUT by Nielsen and Takanishi \ct{37}, having
the group of symmetry $G_{ext}$ (see Eq.(\ref{79y})),
was suggested with aim to explain the neutrino oscillations.
Introducing the right--handed neutrino in the model, the authors replaced the
assumption 1 and considered U(48) group instead of U(45), so that
$G_{ext}$ is a subgroup of U(48): $G_{ext}\subseteq U(48)$. This group
ends up having 7 Higgs fields.

In contrast to the "old" extended AGUT by Froggatt--Nielsen (called here
as $G_f$--theory), the new results of $G_{ext}$--theory are more encouraging,
and it is possible to conclude
that the $G$--theory, in general, is successful in describing of
the SM experiment.

\section{Multiple Point Principle}

AGUT approach is used in conjunction with the Multiple Point
Principle proposed in Ref.\ct{17p}.
According to this principle Nature seeks a special point --- the Multiple
Critical Point (MCP) --- which is a point on the phase diagram of the
fundamental regulirized gauge theory G (or $G_f$, or $G_{ext}$), where
the vacua of all fields existing in Nature are degenerate having the same
vacuum energy density.
Such a phase diagram has axes given by all coupling constants
considered in theory. Then all (or just many) numbers of phases
meet at the MCP.

MPM assumes the existence of MCP at the Planck scale,
insofar as gravity may be "critical" at the Planck scale.

The usual definition of the SM coupling constants:
\begin{equation}
  \alpha_1 = \frac{5}{3}\frac{\alpha}{\cos^2\theta_{\ov{MS}}},\quad
  \alpha_2 = \frac{\alpha}{\sin^2\theta_{\ov{MS}}},\quad
  \alpha_3 \equiv \alpha_s = \frac {g^2_s}{4\pi},     \lb{81y}
\end{equation}
where $\alpha$ and $\alpha_s$ are the electromagnetic and SU(3)
fine structure constants, respectively, is given in the Modified
minimal subtraction scheme ($\ov{MS}$).
Here $\theta_{\ov{MS}}$ is the Weinberg weak angle in $\ov{MS}$ scheme.
Using RGE with experimentally
established parameters, it is possible to extrapolate the experimental
values of three inverse running constants $\alpha_i^{-1}(\mu)$
(here $\mu$ is an energy scale and i=1,2,3 correspond to U(1),
SU(2) and SU(3) groups of the SM) from the Electroweak scale to the Planck
scale. The precision of the LEP data allows to make this extrapolation
with small errors (see \ct{33a}). Assuming that these RGEs for
$\alpha_i^{-1}(\mu)$ contain only the contributions of the SM particles
up to $\mu\approx \mu_{Pl}$ and doing the extrapolation with one
Higgs doublet under the assumption of a "desert", the following results
for the inverses $\alpha_{Y,2,3}^{-1}$ (here $\alpha_Y\equiv \frac{3}{5}
\alpha_1$) were obtained in Ref.\ct{17p} (compare with \ct{33a}):
\begin{equation}
   \alpha_Y^{-1}(\mu_{Pl})\approx 55.5; \quad
   \alpha_2^{-1}(\mu_{Pl})\approx 49.5; \quad
   \alpha_3^{-1}(\mu_{Pl})\approx 54.0.
                                                        \lb{82y}
\end{equation}
The extrapolation of $\alpha_{Y,2,3}^{-1}(\mu)$ up to the point
$\mu=\mu_{Pl}$ is shown in Fig.1.

According to AGUT, at some point $\mu=\mu_G < \mu_{Pl}$ the fundamental
group $G$ (or $G_f$, or $G_{\mbox{ext}}$)
undergoes spontaneous breakdown to its diagonal subgroup:
\begin{equation}
      G \longrightarrow G_{diag.subgr.} = \{g,g,g || g\in SMG\},
                                                          \lb{83y}
\end{equation}
which is identified with the usual (low--energy) group SMG.
The point $\mu_G\sim 10^{18}$ GeV also is shown in Fig.1, together with
a region of G--theory, where AGUT works.

The AGUT prediction of the values of $\alpha_i(\mu)$ at $\mu=\mu_{Pl}$
is based on the MPM assumption about the existence of phase
transition boundary point MCP at the Planck scale, and gives these values
in terms of the corresponding critical couplings $\alpha_{i,crit}$
[\citen{2g},\;\citen{2k},\;\citen{17p}]:
\begin{equation}
            \alpha_i(\mu_{Pl}) = \frac {\alpha_{i,crit}}{N_{gen}}
                       = \frac{\alpha_{i,crit}}{3}
                \quad{\mbox{for}}\quad i=2,3, \,{\mbox{(also for i>3)}},
                                   \lb{84y}
\end{equation}
\begin{equation}
\alpha_1(\mu_{Pl}) = \frac{\alpha_{1,crit}}{\frac{1}{2}N_{gen}(N_{gen} + 1)}
                   = \frac{\alpha_{1,crit}}{6} \quad{\mbox{for}}\quad U(1).
                                      \lb{85y}
\end{equation}

\section{Lattice Theories}

The philosophy of MPM leads to the necessity to investigate the phase
transition in different gauge theories.
A lattice model of gauge theories is the most convenient formalism for the
realization of the MPM ideas. As it was mentioned above, in the simplest
case we can imagine our space--time as a regular hypercubic
(3+1)--lattice with the parameter $a$ equal to the fundamental
(Planck) scale: $a = \lambda_P$.

The lattice SU(N) gauge theories was first introduced by K.Wilson \ct{1s}
for studying the problem of confinement. He suggested the following
simplest action:
\begin{equation}
         S = - \frac{\beta}{N}\sum_p Re(Tr{\cal U}_p),     \lb{36}
\end{equation}
where the sum runs over all plaquettes of a hypercubic lattice
and ${\cal U}_p$ belongs to the fundamental representation of SU(N).
The simplest Wilson lattice action for $U(1)$ gauge theory has the form:
\begin{equation}
     S_W = \beta \sum_p \cos\Theta_p, \quad {\mbox{where}}\quad
                                   {\cal U}_p = e^{i\Theta_p}.   \lb{35a}
\end{equation}
For the compact lattice QED: $\beta = 1/e_0^2$, where $e_0$ is the bare
electric charge.

The Villain lattice action for the $U(1)$ gauge theory is:
\begin{equation}
         S_V = (\beta/2)\sum_p(\Theta_p - 2\pi k)^2, \quad k\in Z.   \lb{39}
\end{equation}

The critical value of the effective electric fine structure
constant $\alpha$
was obtained in Ref.\ct{10s} in the compact QED described by the Wilson and
Villain actions (\ref{35a}) and (\ref{39}), respectively:
\begin{equation}
\alpha_{crit}^{lat}\approx 0.20\pm 0.015\quad
{\mbox{and}} \quad {\tilde \alpha}_{crit}^{lat}\approx 1.25\pm 0.10
\quad{\mbox{at}}\quad
\beta_T\equiv\beta_{crit}\approx{1.011}.
\lb{47}
\end{equation}
Here

\vspace{-12mm}
\begin{equation}
\alpha = \frac{e^2}{4\pi}\quad{\mbox{and}}\quad
\tilde \alpha = \frac{g^2}{4\pi}
\lb{47*}
\end{equation}
are the electric and magnetic fine structure constants, containing
the electric charge $e$ and magnetic charge $g$.

The result of Ref.\ct{10s} for the behavior of $\alpha(\beta)$ in the vicinity
of the phase transition point $\beta_T$ is shown in Fig.2(a) for the Wilson
and Villain lattice actions. Fig.2(b) demonstrates the comparison of the
function $\alpha(\beta)$ obtained by Monte Carlo method for the Wilson
lattice action and by theoretical calculation of the same quantity.
The theoretical (dashed) curve was calculated by so-called "Parisi improvement
formula" \ct{13p}:
\begin{equation}
    \alpha (\beta )=[4\pi \beta W_p]^{-1}.     \lb{48}
\end{equation}
Here $W_p=<\cos \Theta_p >$ is a mean value of the plaquette energy.
The corresponding values of $W_p$ are taken from Ref.\ct{9s}.

The theoretical value of $\alpha_{crit}$ is less than the "experimental"
(Monte Carlo) value (\ref{47}):
\begin{equation}
      \alpha_{crit}\mbox{(in\,\,lattice\,\,theory)}\approx{0.12}.
                                                    \lb{49}
\end{equation}
According to Fig.2(c):
\begin{equation}
   \alpha_{crit.,theor.}^{-1}\approx 8.                   \lb{50a}
\end{equation}
This result does not coincide with the lattice result (\ref{47}),
which gives the following value:
\begin{equation}
   \alpha_{crit.,lat.}^{-1}\approx 5.                   \lb{50b}
\end{equation}
The deviation of theoretical calculations of $\alpha(\beta )$ from the
lattice ones, which is shown in Fig.2(b,c),
has the following explanation: Parisi improvement formula (\ref{48})
is valid in Coulomb--like phase where the mass of artifact monopoles is infinitely
large and photon is massless. But in the vicinity of the phase
transition (critical) point the monopole mass $m\to 0$ and photon
acquires the non--zero mass $m_0\neq 0$ (on side of the confinement). This
phenomenon leads to the "freezing" of $\alpha$: the effective electric fine
structure constant is almost unchanged in the confinement phase and approaches
to its maximal value $\alpha=\alpha_{max}$. The authors of Ref.\ct{14p}
predicted that in the confinement phase, where we have the formation of
strings, the fine structure constant $\alpha$ cannot be infinitely large, but
has the maximal value: $\alpha_{max} \approx \frac{\pi}{12}\approx 0.26$
due to the Casimir effect for strings.

\section{Lattice Artifact Monopoles and Higgs Monopole Model}

Lattice monopoles are responsible for the confinement in lattice
gauge theories what was confirmed by many numerical and theoretical
investigations.

In the previous papers [\citen{17p}--\citen{19p}] the calculations of the U(1)
phase transition (critical) coupling constant were connected with the
existence of artifact monopoles in the lattice gauge theory and also
in the Wilson loop action model \ct{19p}.

In Ref.\ct{19p} we (L.V.L. and H.B.Nielsen) have put forward the speculations
of Refs.[\citen{17p},\citen{18p}] suggesting that the modifications of the form of
the lattice action might not change too much the phase transition value of the
effective continuum coupling constant. The purpose was to investigate this
approximate stability of the critical coupling with respect to a somewhat
new regularization being used instead of the lattice, rather than just
modifying the lattice in various ways.
In \ct{19p} the Wilson loop action was considered in the
approximation of circular loops of radii $R\ge a$. It was shown that the
phase transition coupling constant is indeed approximately independent
of the regularization method: ${\alpha}_{crit}\approx{0.204}$,
in correspondence with the Monte Carlo simulation result on lattice:
${\alpha}_{crit}\approx{0.20\pm 0.015}$ (see Eq.(\ref{47})).

But in Refs.[\citen{20p}--\citen{22p}], instead of using the lattice or Wilson loop
cut--off, we have considered the Higgs Monopole Model (HMM) approximating
the lattice artifact monopoles as fundamental pointlike particles described
by the Higgs scalar field.
The simplest effective dynamics describing the
confinement mechanism in the pure gauge lattice U(1) theory
is the dual Abelian Higgs model of scalar monopoles \ct{13s}, [\citen{20p}--\citen{22p}] (shortly HMM).
This model, first suggested in Refs.\ct{13s}, considers the
following Lagrangian:
\begin{equation}
    L = - \frac{1}{4g^2} F_{\mu\nu}^2(B) + \frac{1}{2} |(\partial_{\mu} -
           iB_{\mu})\Phi|^2 - U(\Phi),\quad              \lb{5y}
{\mbox{where}}\quad
 U(\Phi) = \frac{1}{2}\mu^2 {|\Phi|}^2 + \frac{\lambda}{4}{|\Phi|}^4
\end{equation}
is the Higgs potential of scalar monopoles with magnetic charge $g$, and
$B_{\mu}$ is the dual gauge (photon) field interacting with the scalar
monopole field $\Phi$.  In this model $\lambda$ is the self--interaction
constant of scalar fields, and the mass parameter $\mu^2$ is negative.
In Eq.(\ref{5y}) the complex scalar field $\Phi$ contains
the Higgs ($\phi$) and Goldstone ($\chi$) boson fields:
\begin{equation}
          \Phi = \phi + i\chi.             \lb{7y}
\end{equation}
The effective potential in the Higgs Scalar ElectroDynamics (HSED)
was first calculated by Coleman and Weinberg \ct{20s} in the one--loop
approximation. The general method of its calculation is given in the
review \ct{21s}. Using this method, we can construct the effective potential
for HMM. In this case the total field system of the gauge ($B_{\mu}$)
and magnetically charged ($\Phi$) fields is described by
the partition function which has the following form in Euclidean space:
\begin{equation}
      Z = \int [DB][D\Phi][D\Phi^{+}]\,e^{-S},     \lb{8y}
\end{equation}
where the action $S = \int d^4x L(x) + S_{gf}$ contains the Lagrangian
(\ref{5y}) written in Euclidean space and gauge fixing action $S_{gf}$.
Let us consider now a shift:
\begin{equation}
 \Phi (x) = \Phi_b + {\hat \Phi}(x)                \lb{9y}
\end{equation}
with $\Phi_b$ as a background field and calculate the
following expression for the partition function in the one-loop
approximation:
$$
  Z = \int [DB][D\hat \Phi][D{\hat \Phi}^{+}]
   \exp\{ - S(B,\Phi_b)
   - \int d^4x [\frac{\delta S(\Phi)}{\delta \Phi(x)}|_{\Phi=
   \Phi_b}{\hat \Phi}(x) + h.c. ]\}\\
$$
\begin{equation}
    =\exp\{ - F(\Phi_b, g^2, \mu^2, \lambda)\}.      \lb{10y}
\end{equation}
Using the representation (\ref{7y}), we obtain the effective potential:
\begin{equation}
  V_{eff} = F(\phi_b, g^2, \mu^2, \lambda)        \lb{11y}
\end{equation}
given by the function $F$ of Eq.(\ref{10y}) for the constant background
field $ \Phi_b = \phi_b = \mbox{const}$. In this case the one--loop
effective potential for monopoles coincides with the expression of the
effective potential calculated by the authors of Ref.\ct{20s} for scalar
electrodynamics and extended to the massive theory (see review \ct{21s}).

Considering the renormalization group improvement
of the effective Coleman--Weinberg potential \ct{20s},\ct{21s}, written
in Ref.\ct{22p} for the dual sector of scalar electrodynamics in the
two--loop approximation, we have calculated the U(1) critical values of
the magnetic fine structure constant:
\begin{equation}
{\tilde\alpha}_{crit} = g^2_{crit}/4\pi\approx 1.20,      \lb{1cr}
\end{equation}
and electric fine structure constant:
\begin{equation}
 \alpha_{crit} = \pi/g^2_{crit}\approx 0.208             \lb{2cr}
\end{equation}
by the Dirac relation:
\begin{equation}
          eg= 2\pi, \quad{\mbox{or}}\quad \alpha\tilde \alpha = \frac{1}{4}.
                                                       \lb{3dr}
\end{equation}
The values (\ref{1cr}),(\ref{2cr}) coincide with the lattice result
(\ref{47}).

\section{Monopoles strength group dependence}

As it was shown in a number of investigations, the confinement in the SU(N) lattice gauge
theories effectively comes to the same U(1) formalism. The reason is the
Abelian dominance in their monopole vacuum: monopoles of the Yang--Mills
theory are the solutions of the U(1)--subgroups, arbitrary embedded into
the SU(N) group. After a partial gauge fixing (Abelian projection by
't Hooft \ct{24p}) SU(N) gauge theory is reduced to an Abelian
$U(1)^{N-1}$ theory with $N-1$ different types of Abelian monopoles.
Choosing the Abelian gauge for dual gluons, it is possible to describe
the confinement in the lattice SU(N) gauge theories by the analogous
dual Abelian Higgs model of scalar monopoles.

Considering the Abelian gauge and taking into account that
the direction in the Lie algebra of monopole fields are gauge
independent, we have found in Ref.\ct{22p} an average over these directions
and obtained \un{the group dependence relation} between the phase transition
fine structure constants for the groups $U(1)$ and $SU(N)/Z_N$:
\begin{equation}
      \alpha_{N,crit}^{-1}
           = \frac{N}{2}\sqrt{\frac{N+1}{N-1}}
                          \alpha_{U(1),crit}^{-1}.
                                            \lb{25z}
\end{equation}

\section{AGUT-MPM prediction of the Planck scale values of the
U(1), SU(2) and SU(3) fine structure constants}

As it was assumed in Ref.\ct{17p}, the MCP values
$\alpha_{i,crit}$ in Eqs.(\ref{84y}) and (\ref{85y}) coincide with
the critical values of the effective fine structure
constants given by the generalized lattice SU(3)--, SU(2)-- and U(1)--gauge
theories.

Now let us consider $\alpha_Y^{-1}\,(\approx \alpha^{-1})$ at the point
$\mu=\mu_G\sim 10^{18}$ GeV shown in Fig.1.
If the point $\mu=\mu_G$ is very close to the Planck scale
$\mu=\mu_{Pl}$, then according to Eqs.(\ref{82y}) and (\ref{85y}), we have:
\begin{equation}
         \alpha_{1st\, gen.}^{-1}\approx
    \alpha_{2nd\, gen.}^{-1}\approx \alpha_{3rd\, gen.}^{-1}\approx
    \frac{\alpha_Y^{-1}(\mu_G)}{6}\approx 9,        \lb{88y}
\end{equation}
what is almost equal to the value (\ref{50a}):

\vspace*{-12.5mm}
$$
            \alpha_{crit.,theor}^{-1}\approx 8
$$
obtained by the Parisi improvement method (see Fig.2(c)).
This means that in the U(1) sector of AGUT we have $\alpha $ near
the critical point. Therefore, we can expect the existence of MCP
at the Planck scale.

It is necessary to mention that the lattice investigators were not able
to obtain the lattice triple point values $\alpha_{i,crit}$
(i=1,2,3 correspond to U(1),SU(2) and SU(3) groups) by Monte Carlo
simulation methods.
These values were calculated theoretically by Bennett and Nielsen
in Ref.\ct{17p} using the Parisi improvement method \ct{13p}:
\begin{equation}
    \alpha_{Y,crit}^{-1}\approx 9.2\pm 1,
    \quad \alpha_{2,crit}^{-1}\approx 16.5\pm 1, \quad
    \alpha_{3,crit}^{-1}\approx 18.9\pm 1.                 \lb{89y}
\end{equation}
Assuming the existence of MCP at $\mu=\mu_{Pl}$
and substituting the last results in Eqs.(\ref{84y}) and (\ref{85y}),
we have the following prediction of AGUT \ct{17p}:
\begin{equation}
   \alpha_Y^{-1}(\mu_{Pl})\approx 55\pm 6; \quad
   \alpha_2^{-1}(\mu_{Pl})\approx 49.5\pm 3; \quad
   \alpha_3^{-1}(\mu_{Pl})\approx 57.0\pm 3.
                                                          \lb{90y}
\end{equation}
These results coincide with the results (\ref{82y}) obtained by the
extrapolation of experimental data to the Planck scale
in the framework of pure SM (without any new particles) \ct{33a}, \ct{17p}.

Using the relation (\ref{25z}), we obtained the following relations:
\begin{equation}
    \alpha_{Y,crit}^{-1} : \alpha_{2,crit}^{-1} : \alpha_{3,crit}^{-1}
           = 1 : \sqrt{3} : 3/\sqrt{2} = 1 : 1.73 : 2.12.
                                                     \lb{91y}
\end{equation}
Let us compare now these relations with the MPM prediction.

For $\alpha_{Y,crit}^{-1}\approx 9.2$  given by the first equation of
(\ref{89y}), we have:
\begin{equation}
 \alpha_{Y,crit}^{-1} : \alpha_{2,crit}^{-1} : \alpha_{3,crit}^{-1}
    = 9.2 : 15.9 : 19.5.                                     \lb{92y}
\end{equation}
In the framework of errors the last result coincides with the
AGUT--MPM prediction (\ref{89y}).
Of course, it is necessary to take into account an approximate description
of confinement dynamics in the SU(N) gauge theories, which was
used in our investigations.

\section{The possibility of the Grand Unification Near the Planck Scale}

We can see new consequences of the extension of $G$--theory, if
$G$--group is broken down to its diagonal subgroup $G_{diag}$, i.e. SM,
not at $\mu_G\sim 10^{18}$ {GeV}, but at $\mu_G\sim 10^{15}$ {GeV}.
In this connection, it is very attractive to consider the gravitational
interaction.

\subsection{"Gravitational finestructure constant" evolution}

The gravitational interaction between two particles
of equal masses M is given by the usual classical Newtonian potential:
\begin{equation}
   V_g = - G \frac{M^2}{r} =
           - \left(\frac{M}{M_{Pl}}\right)^2\frac{1}{r}
                   = - \frac{\alpha_g(M)}{r},              \lb{1x}
\end{equation}
which always can be imagined as a tree--level approximation of quantum
gravity.

Then the quantity:
\begin{equation}
      \alpha_g = \left(\frac{\mu}{\mu_{Pl}}\right)^2     \lb{2x}
\end{equation}
plays a role of the "gravitational finestructure constant" and the
evolution of its inverse quantity is presented in Fig.3 together with the
evolutions of $\alpha_{1,2,3}^{-1}(\mu)$ (here we have returned to the
consideration of $\alpha_1$ instead of $\alpha_Y$).

Then we see the intersection of $\alpha_g^{-1}(\mu)$
with $\alpha_1^{-1}(\mu)$ in the region of $G$--theory at the point:
$$
               (x_0, \alpha_0^{-1}),
$$
where
\begin{equation}
      x_0 \approx 18.3,  \quad {\mbox{and}}\quad
       \alpha_0^{-1} \approx 34.4,                   \lb{3x}
\end{equation}
and $\,x = \log_{10}\mu$.

\subsection{The consequences of the breakdown of $G$-theory
at $\mu_G\sim 10^{15}$ or $10^{16}$ GeV}

Let us assume now that the group of symmetry $G$ undergoes the breakdown
to its diagonal subgroup not at $\mu_G\sim 10^{18}$ GeV, but at
$\mu_G\sim 10^{15}$ GeV, i.e. before the intersection of
$\alpha_{2}^{-1}(\mu)$ and $\alpha_{3}^{-1}(\mu)$ at
$\mu\sim 10^{16}$
GeV.

As a consequence of behavior of the function $\alpha^{-1}(\beta)$
near the phase transition point, shown in Fig.2c, we have to expect the
change of the evolution of $\alpha_i^{-1}(\mu)$ in the region $\mu > \mu_G$
shown in Fig.1 by dashed lines. Instead of these dashed lines,
we must see the decreasing of $\alpha_i^{-1}(\mu)$, when they
approach MCP, if this MCP really exists at the Planck scale.

According to Fig.2c, in the very vicinity of the phase transition point
(i.e. also near the MCP at $\mu=\mu_{Pl}$), we cannot
describe the behavior of $\alpha_i^{-1}(\mu)$ by the one---loop
approximation RGE.

It is well known, that the one--loop approximation RGEs
for $\alpha_i^{-1}(\mu)$ can be
described in our case by the following expression~\ct{1a}:
\begin{equation}
  \alpha_i^{-1}(\mu) =
  \alpha_i^{-1}(\mu_{Pl}) + \frac{b_i}{4\pi}\log(\frac{\mu^2}{\mu^2_{Pl}}),
                                                \lb{4x}
\end{equation}
where $b_i$ are given by the following values:
$$
   b_i = (b_1, b_2, b_3) =
$$
\begin{equation}
( - \frac{4N_{gen}}{3} -\frac{1}{10}N_S,\,\,
      \frac{22}{3}N_V - \frac{4N_{gen}}{3} -\frac{1}{6}N_S,\,\,
      11 N_V - \frac{4N_{gen}}{3} ).                   \lb{5x}
\end{equation}
The integers $N_{gen},\,N_S,\,N_V\,$ are respectively the numbers
of generations,
Higgs bosons and different vector gauge fields of given "colors".

For the SM we have:
\begin{equation}
       N_{gen} = 3, \quad N_S = N_V =1,                    \lb{6x}
\end{equation}
and the corresponding slopes describe the evolutions of
$\alpha_i^{-1}(\mu)$ up to $\mu = \mu_G$ presented in Fig.3.

But in the region $\mu_G\le \mu \le \mu_{Pl}$, when $G$--theory works,
we have $N_V = 3$ (here we didn't take into account the additional
Higgs fields which can change the number $N_S$), and the one--loop
approximation slopes are almost 3 times larger than the same ones for the SM.
In this case, it is difficult to understand that such evolutions give the
MCP values of $\alpha_i^{-1}(\mu_{Pl})$, which are
shown in Fig.4. These values were obtained by the following way:
$$
  \alpha_1^{-1}(\mu_{Pl}) \approx
  6\cdot \frac{3}{5}\alpha_{U(1),crit}^{-1}\approx 13, \quad
$$
$$
  \alpha_2^{-1}(\mu_{Pl}) \approx
  3\cdot \sqrt{3}\alpha_{U(1),crit}^{-1}\approx 19,\quad
$$
\begin{equation}
  \alpha_3^{-1}(\mu_{Pl}) \approx
  3\cdot \frac{3}{\sqrt 2}\alpha_{U(1),crit}^{-1}\approx 24,      \lb{7x}
\end{equation}
where we have used the relation (\ref{25z}) with
\begin{equation}
  \alpha_{U(1),crit} =
  \frac{\alpha_{crit}}{\cos^2\theta_{\ov{MS}}}\approx 0.77\alpha_{crit},
                      \lb{8x}
\end{equation}
taking into account our HMM result (\ref{2cr}):
$\alpha_{crit}\approx 0.208$, which coincides with the lattice result
(\ref{47}) and gives:
\begin{equation}
  \alpha_{U(1),crit}^{-1} \approx 3.7.
                                            \lb{9x}
\end{equation}
In the case when $G$--group undergoes the breakdown to the SM not
at $\mu_G\sim 10^{18}$ GeV, but at $\mu_G\sim 10^{15}$ GeV, the artifact
monopoles of non-Abelian  SU(2) and SU(3) sectors of $G$--theory begin
to act more essentially.

According to the group dependence relation (\ref{25z})
(although now it is necessary to expect that it is very approximate)
we have, for example, the following estimation
at $\mu_G\sim 10^{15}$ GeV:
\begin{equation}
   \alpha_{U(1)}^{-1}(\mu_G) \sim 7\quad -\quad for\quad
       SU(3)_{1st\,\,gen.}, \,etc.                         \lb{10x}
\end{equation}
which is closer to MCP than the previous value $\alpha_Y^{-1}\sim 9$,
obtained for the AGUT breakdown at $\mu_G\sim 10^{18}$ GeV.

It is possible to assume that $\beta$--functions of SU(2) and SU(3)
sectors of $G$--theory change their one--loop approximation behavior
in the region $\mu > 10^{16}$ GeV and $\alpha_{2,3}^{-1}(\mu)$
begin to decrease, approaching the phase transition (multiple critical)
point at $\mu = \mu_{Pl}$. This means that the asymptotic freedom
of non--Abelian theories becomes weaker near the Planck scale, what can
be explained by the influence of artifact monopoles.
It looks as if these $\beta$--functions have
singularity at the phase transition point and, for example,
can be approximated by the following expression:
\begin{equation}
   \frac{d\alpha^{-1}}{dt} = \frac{\beta(\alpha)}{\alpha}\approx
             A(1 - \frac{\alpha}{\alpha_{crit}})^{-\nu}\quad
              {\mbox{near the phase transition point}}.       \lb{11x}
\end{equation}
This possibility is shown in Fig.4 for $\nu\approx 1$ and $\nu \approx 2.4$.

Here it is worth-while to comment that such a tendency was revealed
in the vicinity of the confinement phase by the forth--loop
approximation of $\beta$--function in QCD (see Ref.\ct{40a}).

\subsection{Does the [SU(5)]$^{\bf 3}$ SUSY unification exist near the Planck
scale?}

Approaching the MCP in the region of $G$--theory ($\mu_G\le \mu \le \mu_{Pl}$),
$\alpha_{2,3}^{-1}(\mu)$ show the necessity
of intersection of $\alpha_{2}^{-1}(\mu)$ with $\alpha_{3}^{-1}(\mu)$
at some point of this region if $\mu_G\sim 10^{15}$ or $10^{16}$ GeV
(see Fig.4).
If this intersection takes place at the point $(x_0,\,\alpha_0^{-1})$
given by Eq.(\ref{3x}), then we have the unification of all gauge
interactions (including the gravity) at the point:
\begin{equation}
  (x_{GUT};\,\alpha_{GUT}^{-1})\approx (18.3;\,34.4),     \lb{12x}
\end{equation}
where $x = \log_{10}\mu$(GeV).
Here we assume the existence of [SU(5)]$^3$ SUSY unification having superparticles of masses
\begin{equation}
           M\approx 10^{18.3}\, {\mbox{GeV}}.        \lb{13x}
\end{equation}
The scale $\mu_{GUT}=M$, given by Eq.(\ref{13x}), can be considered
as a SUSY breaking scale.

Figures 5(a,b) demonstrates such a possibility of unification.
We have investigated the solutions of joint intersections of
$\alpha_g^{-1}(\mu)$ and all
$\alpha_i^{-1}(\mu)$ at different $x_{GUT}$ with different $\nu$
in Eq.(\ref{11x}). These solutions exist from
$\nu \approx 0.5$ to $\nu \approx 2.5$.

The unification theory with [SU(5)]$^3$--symmetry was suggested first
by S.Rajpoot \ct{40}.

It is essential that the critical point in this theory,
obtained by means of Eqs.(\ref{84y}), (\ref{25z}) and (\ref{9x}),
is given by the following value:
\begin{equation}
      \alpha_{5,crit}^{-1}\approx 3\cdot \frac{5}{2}\sqrt
                              {\frac{3}{2}}\approx 34.0.      \lb{14x}
\end{equation}
The point (\ref{14x}) is shown in Fig.5(a,b) presented for the cases:

1. $\nu \approx 1$, $\alpha^{-1}_{GUT}\approx 34.4$, $x_{GUT}\approx 18.3$
and $\mu_G\approx 10^{16}$ GeV shown in Fig.5(a);

2. $\nu \approx 2.4$, $\alpha^{-1}_{GUT}\approx 34.4$, $x_{GUT}\approx 18.3$
and $\mu_G\approx 10^{15}$ GeV shown in Fig.5(b).

We see that the point (\ref{14x}) is very close to the
unification point $\alpha_{GUT}^{-1}\approx 34.4$ given by Eq.(\ref{12x}).
This means that the unified theory, suggested here as the [SU(5)]$^3$ SUSY
unification, approaches the confinement phase at the Planck scale.
But the confinement of all SM particles is impossible in our world,
because then they have to be confined also at low energies what is not
observed in the Nature.

It is worth--while to mention that using the Zwanziger formalism for the
Abelian gauge theory with electric and magnetic charges
(see Refs.[\citen{41}--\citen{43}] and \ct{21p}), the possibility of unification
of all gauge interactions at the Planck scale was considered in Ref.\ct{38}
in the case when unconfined monopoles come to the existence near the Planck
scale. They can appear only in G--theory, because RGEs for monopoles
strongly forbid their deconfinement in the SM up to the Planck scale.
But it is not obvious that they really exist in the G--theory. This problem
needs more careful investigations, because our today knowledge about
monopoles is still very poor.

The unified theory, suggested in this paper, essentially differs in its
origin from the case considered in Ref.\ct{38}, because this theory
does not assume the existence of deconfining monopoles up to the
Planck scale, but assumes the influence of our space--time lattice
artifact monopoles near the phase transition (critical) point.

Considering the predictions of this unified theory for the low--energy
physics and cosmology, maybe in future we shall be able to answer
the question: "Does the [SU(5)]$^3$ SUSY unification
theory really exist near the Planck scale?"

\bc
\noindent\includegraphics[width=100mm, height=85mm]{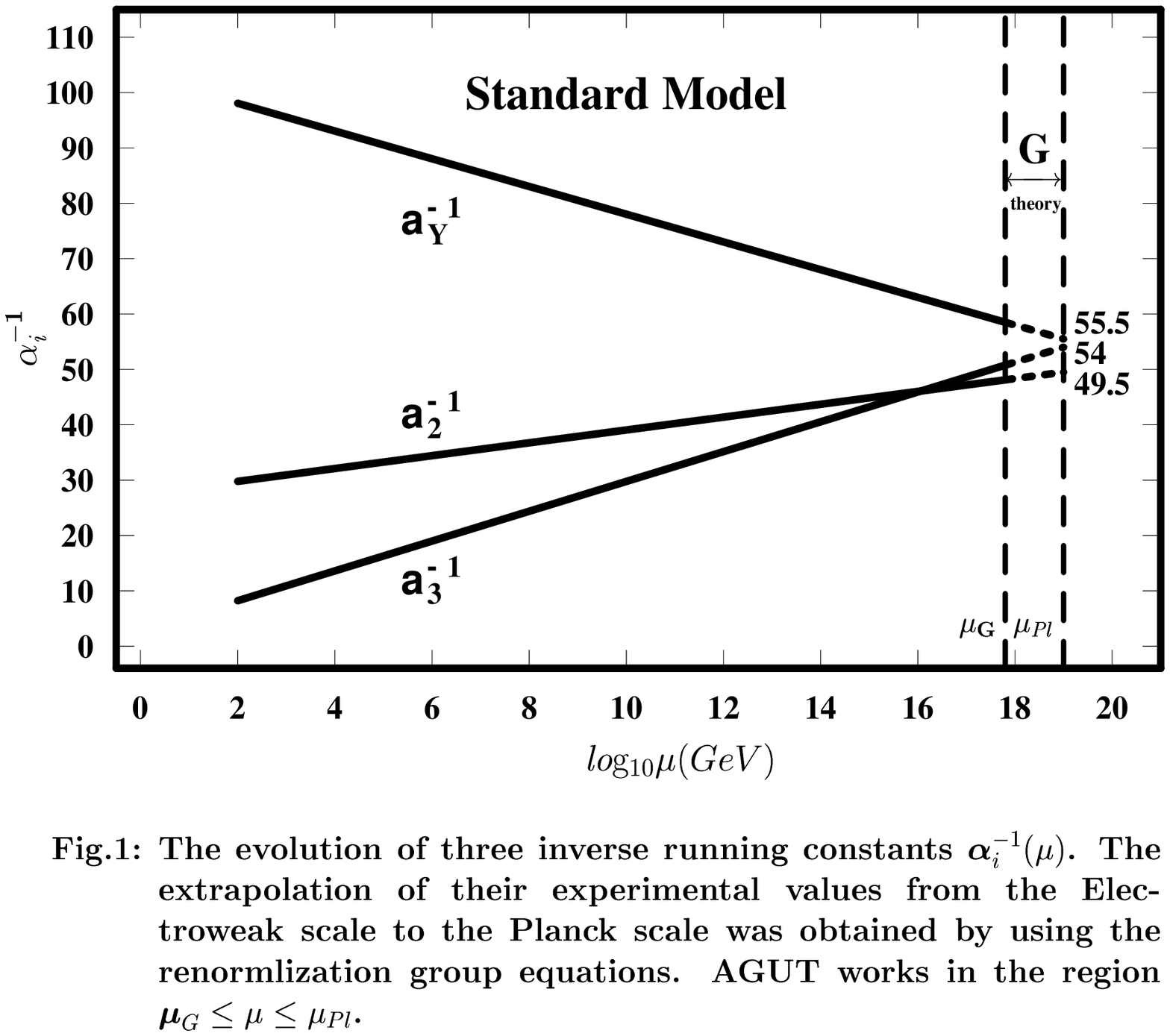}

\vspace*{3mm}
\noindent\includegraphics[width=100mm, height=142mm]{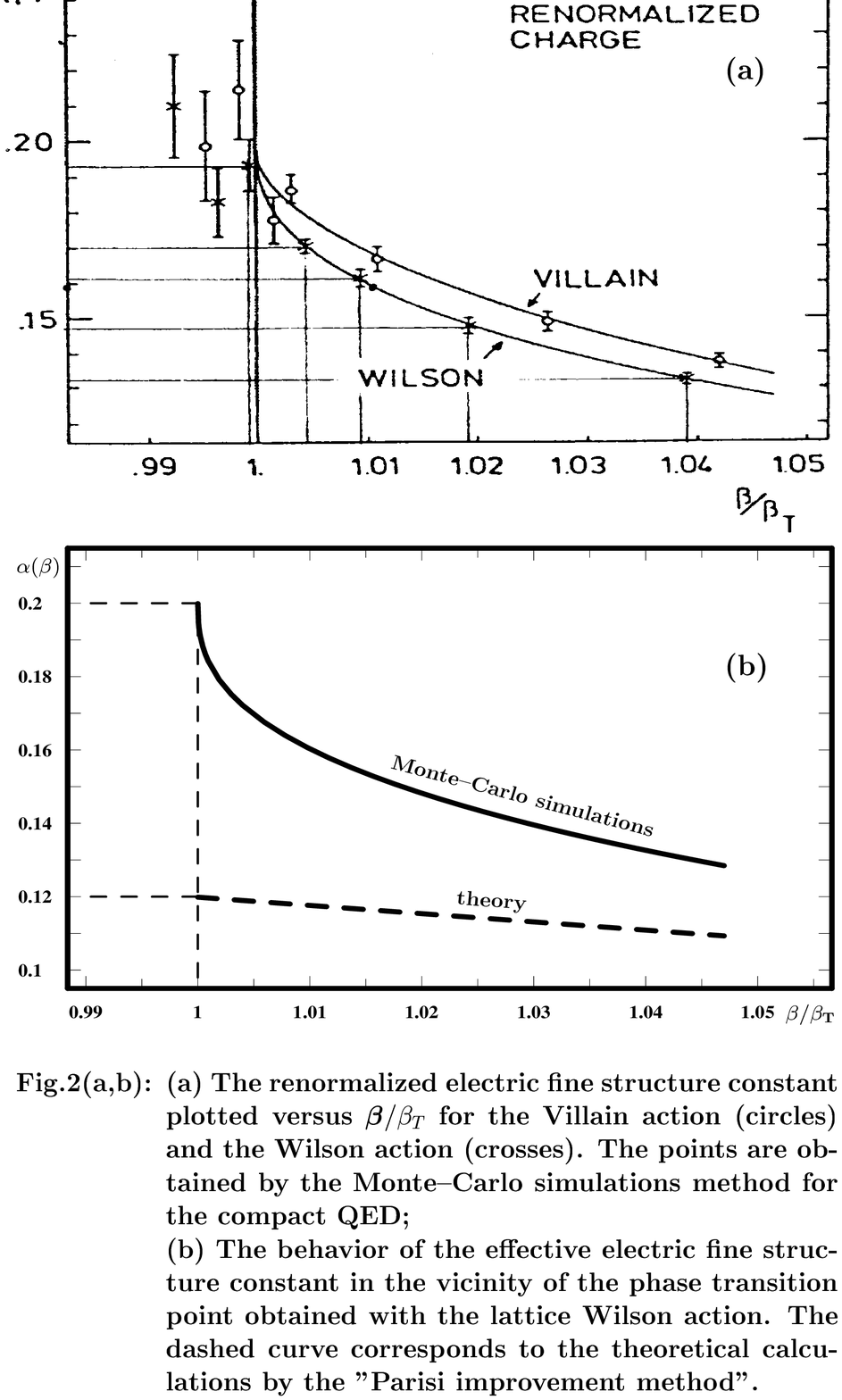}
\ec

\bc
\noindent\hspace*{-5mm}\includegraphics[width=130mm, height=105mm]{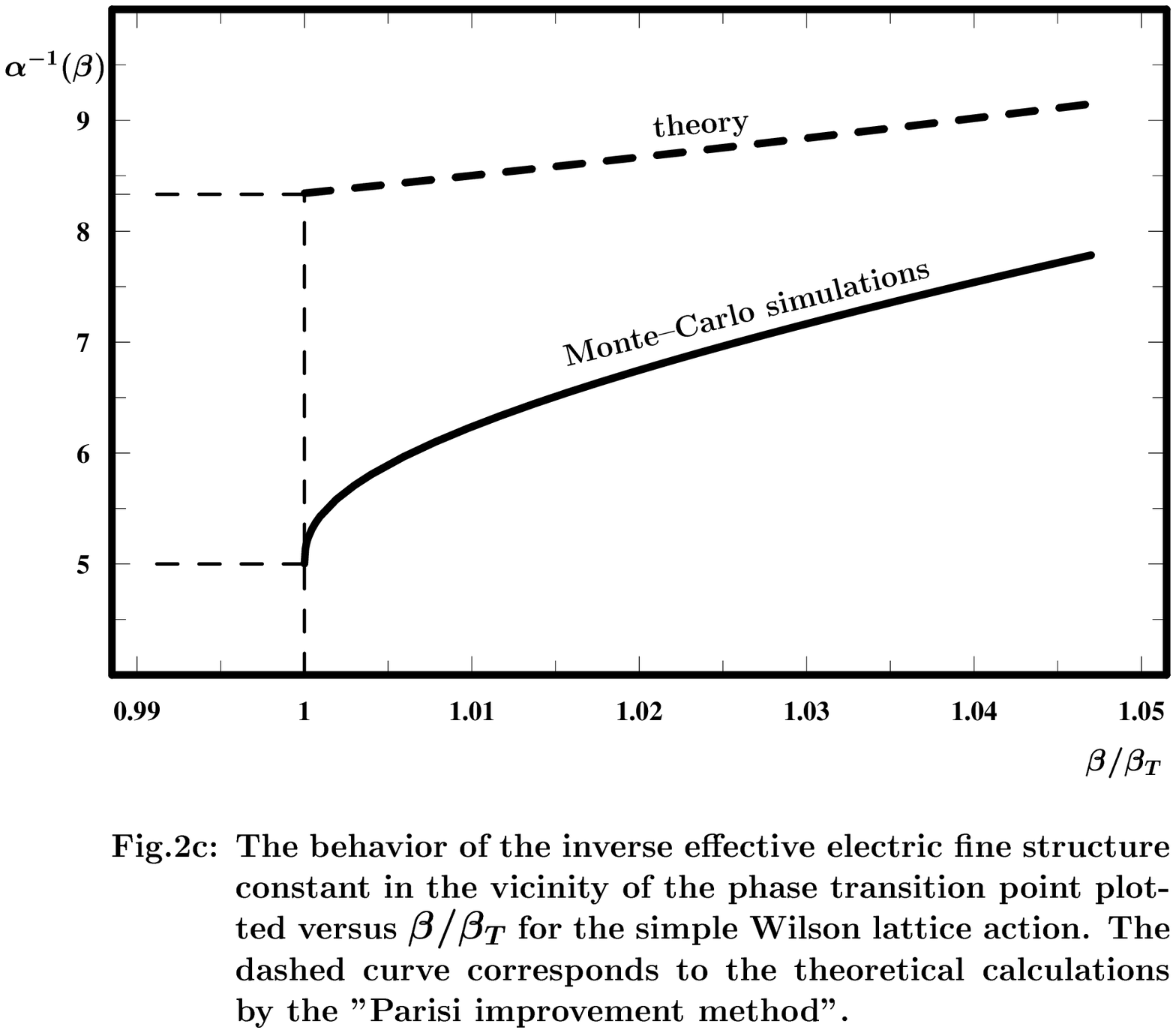}

\vspace*{5mm}
\noindent\hspace*{-5mm}\includegraphics[width=130mm, height=110mm]{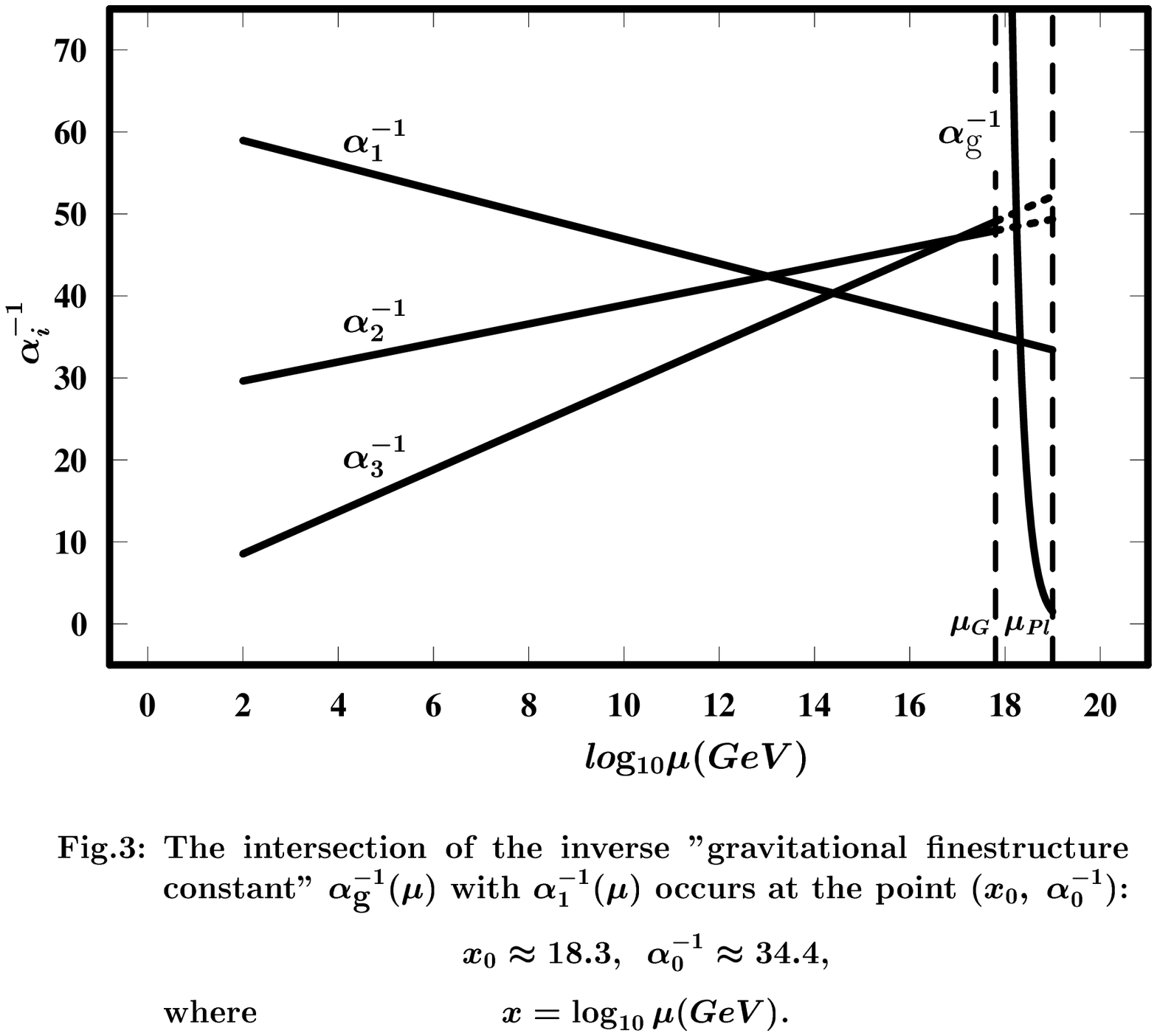}
\ec

\bc
\noindent\hspace*{-5mm}\includegraphics[width=85mm, height=113mm]{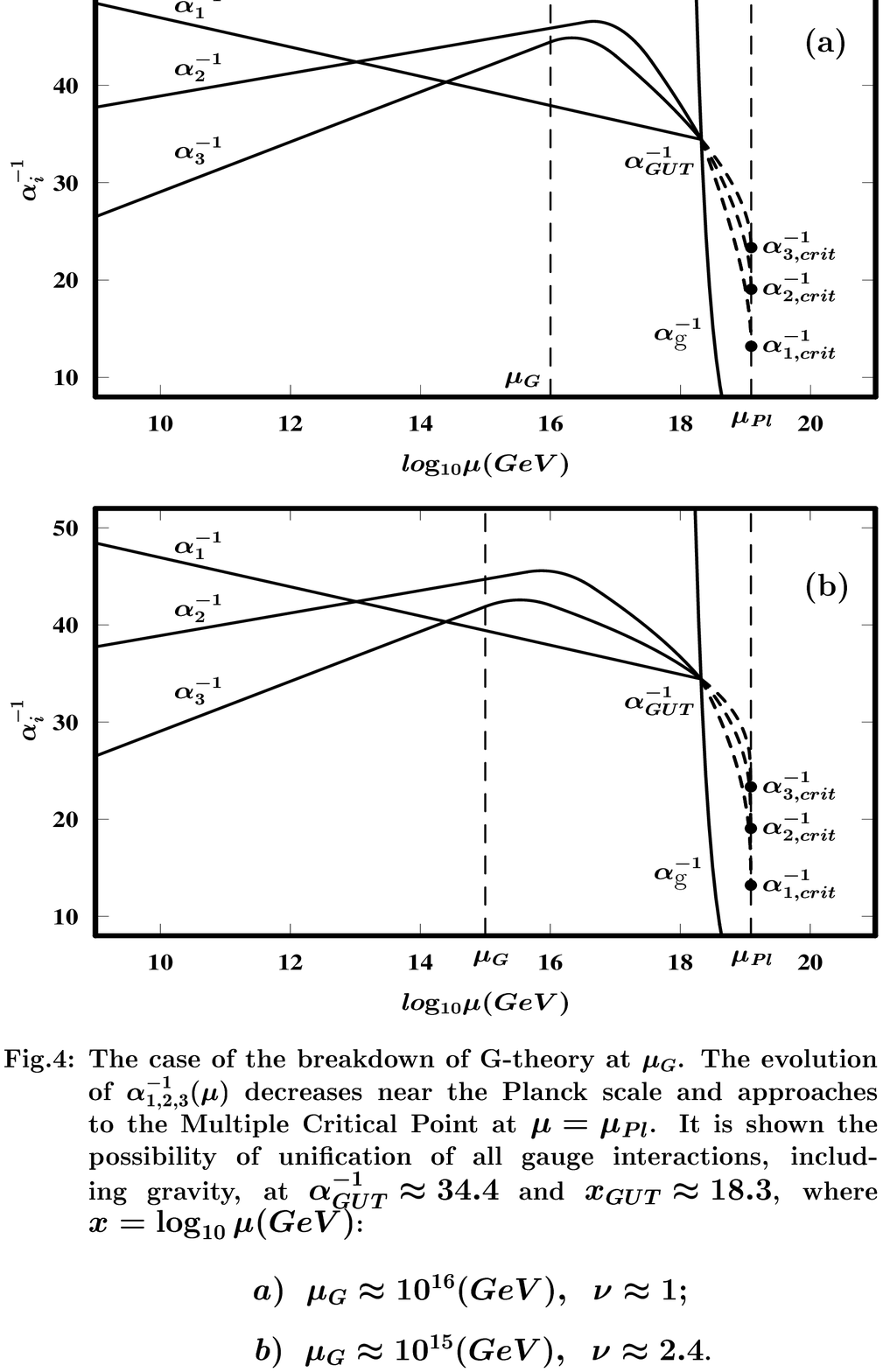}

\vspace*{3mm}
\noindent\hspace*{-5mm}\includegraphics[width=85mm, height=113mm]{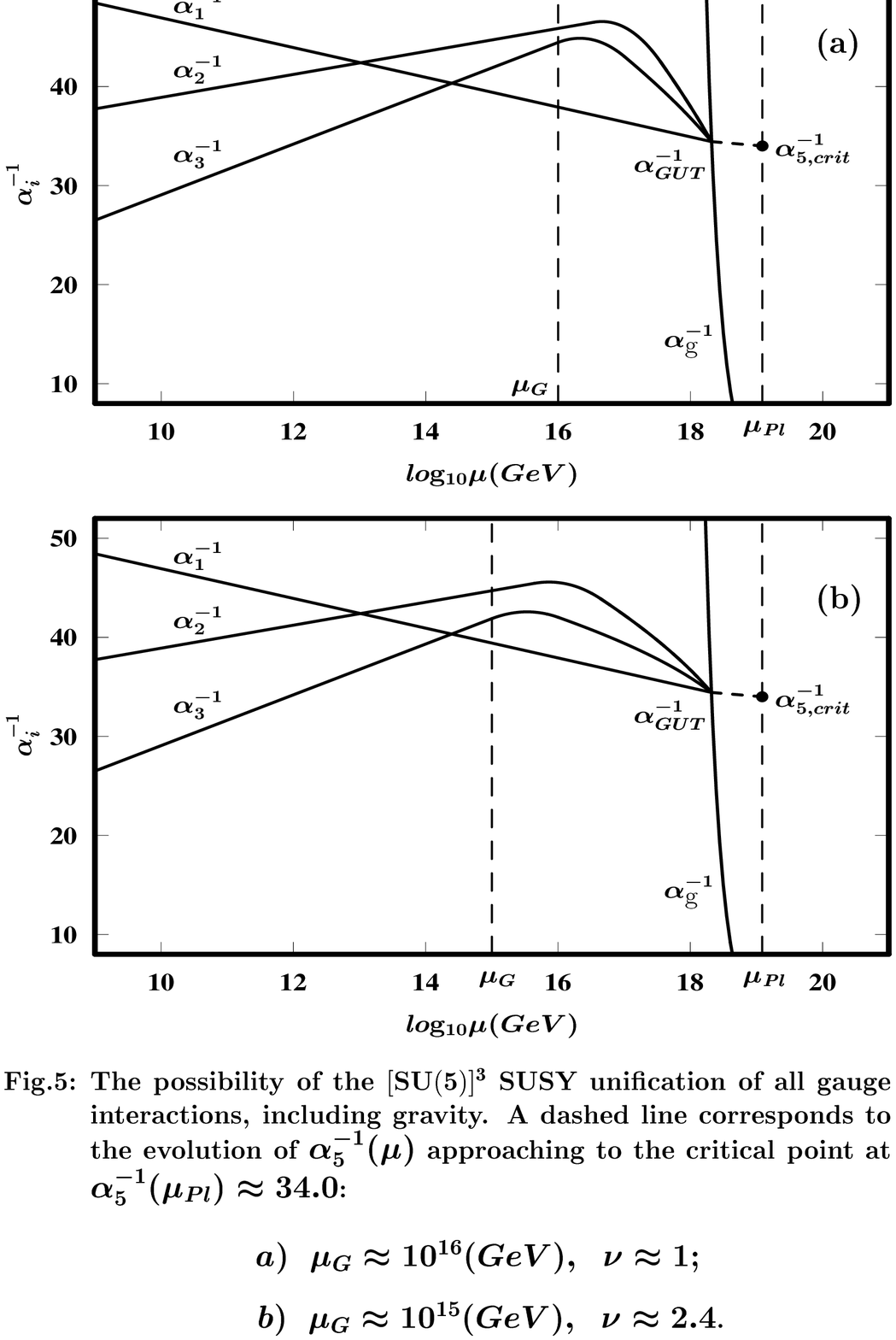}
\ec


\end{document}